# Physisorption on Nanomechanical Resonators: The Overlooked Influence of Trace Moisture

Hemant Kumar Verma[1], Suman Kumar Mandal[1], Darkasha Khan[1], Faizan Tariq Beigh[1], Manoj Kandpal[2], Jaspreet Singh[2], Sushobhan Avasthi[1], Srinivasan Raghavan[1] and Akshay Naik[1*]

[1]Centre for Nano Science and Engineering, Indian Institute of Science, Bangalore 560012, India.
[2]Semi-Conductor Laboratory, Mohali 160071, India.

*anaik@iisc.ac.in

## ABSTRACT

Short gas pulses introduced in a vacuum chamber have long been utilized to showcase the ultra-low mass resolutions achievable with nanomechanical resonators. The resonance frequency shifts are used as evidence of gas adsorption. However, there is very little clarity as to what exactly is adsorbing on to the resonators. We demonstrate that the physisorption of gases on cantilevers is predominantly the effect of moisture content that is present even in ultra-high purity gases. The experimental work is performed at low temperatures and in a high vacuum and is supported by theoretical calculations and simulation.

## INTRODUCTION

Physisorption plays a critical role in important processes like material growth [1], catalysis [2], micropore analysis [3], and adhesion [4]. In mechanical resonators, physisorption affects damping [5] and noise [6]. More importantly, the physisorption of gases has been used to quantify the mass resolution of nanomechanical resonators [7-10]. Physisorption-based gas sensing relies on introducing ultrahigh purity (UHP) gas pulses into a vacuum chamber with a cooled mechanical resonator. The resulting frequency shift is attributed to gas adsorption. However, no direct evidence confirms that the adsorbed species is the intended gas, rather than an impurity such as moisture. Ultra-high pure gases often contain a few ppm of impurities, including moisture, which could also cause the shift in the resonator. Goal of this work is to identify whether the adsorbed gas on the resonators is the target gas, moisture, or a combination of the two.

In this work, we provide experimental evidence that frequency shifts in cooled microcantilevers are largely due to trace moisture in the vacuum chamber and not the target gas. The experimental setup is similar to previous experiments, short pulses of ultrahigh pure gas are introduced into a vacuum chamber with cooled micro resonators [11,12]. However, we also measure and model the temperature-programmed temporal variations in the microcantilevers, revealing that the frequency shift is due to the moisture adsorption. The results show that gas sensors involving physisorption must carefully evaluate the impact of residual gases on the frequency shifts. The physical phenomenon originating in the physisorption processes, like friction and frequency noise, must be reevaluated as they may also be side-effects of moisture [5,6].

**EXPERIMENTAL SECTION**

We use an uncoated silicon nitride microcantilever as the resonator in these experiments. Figure 1(a) is the cross-sectional schematic of the surface micromachined microcantilever. These devices were fabricated on a highly-doped silicon, also serving as the back gate for electrostatic actuation. The structural layer comprises a ~300 nm thick silicon nitride film. Microcantilever's motion is detected using metal piezo resistors (Cr/Au with a thickness of ~ 5 nm/25 nm) in a Wheatstone-bridge configuration [13][14].

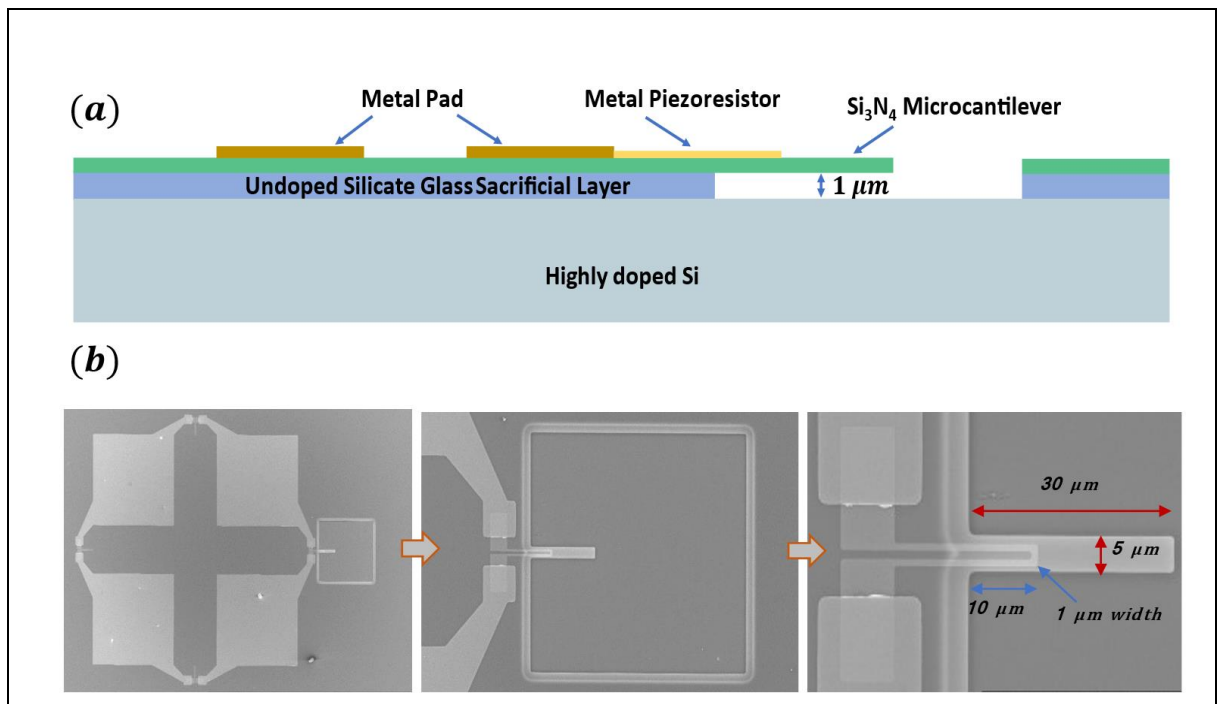

*Figure 1 : (a) Diagram illustrating the cross-sectional view of a silicon nitride microcantilever. The microcantilever has dimensions of 30 μm length, 5 μm width, and a thickness of approximately 300 nm. (b) Electron micrograph of the microcantilever device showing one piezoresistor positioned on top of the cantilever while the others are on the fixed substrate.*

The microcantilever is wire-bonded to a read-out printed circuit board (PCB) mounted on a cryostat (Advance Research System). The piezoresistive signal is amplified using a voltage pre-amplifier (SRS 560) and fed into a lock-in amplifier (UHFLI from Zurich Instruments). The details of the measurement circuit are in the Supplementary Information (SI) section 2. Actuation voltage was selected to ensure a linear device response.

The uncoated microcantilever was exposed to ultra-high purity (UHP) gases in a vacuum chamber with a base pressure of $10^{-5}$ mbar (Figure S3 SI). The UHP gases are connected to the vacuum chamber. The quantity of UHP gases introduced into the gas chamber is controlled using two ball valves. The dosage is defined by the volume between the two valves. The resulting frequency response of the device is tracked using a lock-in amplifier with a phase-lock loop (PLL).

Figure. 2(a) shows a graphical representation of the expected adsorption-desorption dynamics of gases at various times as the gas pulse enters the system and is subsequently pumped. Each time the gas pulse enters the chamber, the pressure rises to $1.7 \times 10^{-2}$ mbar before returning to its original level of $3.4 \times 10^{-5}$ mbar, and we observe a frequency shift. The implicit assumption in previous physio adsorption-based gas sensing experiments is that the target gas is adsorbing onto the cantilever [11,12]. However, even ultra-high purity gases have trace amounts of impurities, including moisture. Some literature suggests that moisture in the gas can also physically adsorb onto the silicon nitride microcantilever surface through hydrogen bonding [15,16]. We have quantified the moisture content in UHP gas using the cavity ring-down spectroscopy (CRDS) from HALO Tiger Optics. The measured moisture in UHP helium and argon was ~18 parts per million (ppm) and ~9 ppm, respectively (Figure S4 SI). The adsorption and desorption processes are significantly influenced by the temperature of the adsorbent and the ambient pressure [17]. However, in this study, we focused on temperature programmed adsorption-desorption phenomena keeping the chamber pressure constant for the gases. The interaction between inert gases and silicon nitride surface is relatively weak and mediated through van der Waals forces. [18].

Figures 2(b)-(e) show the temperature-dependent responses of the uncoated microcantilever when pulses of helium gas are flowed into the chamber. We observe a

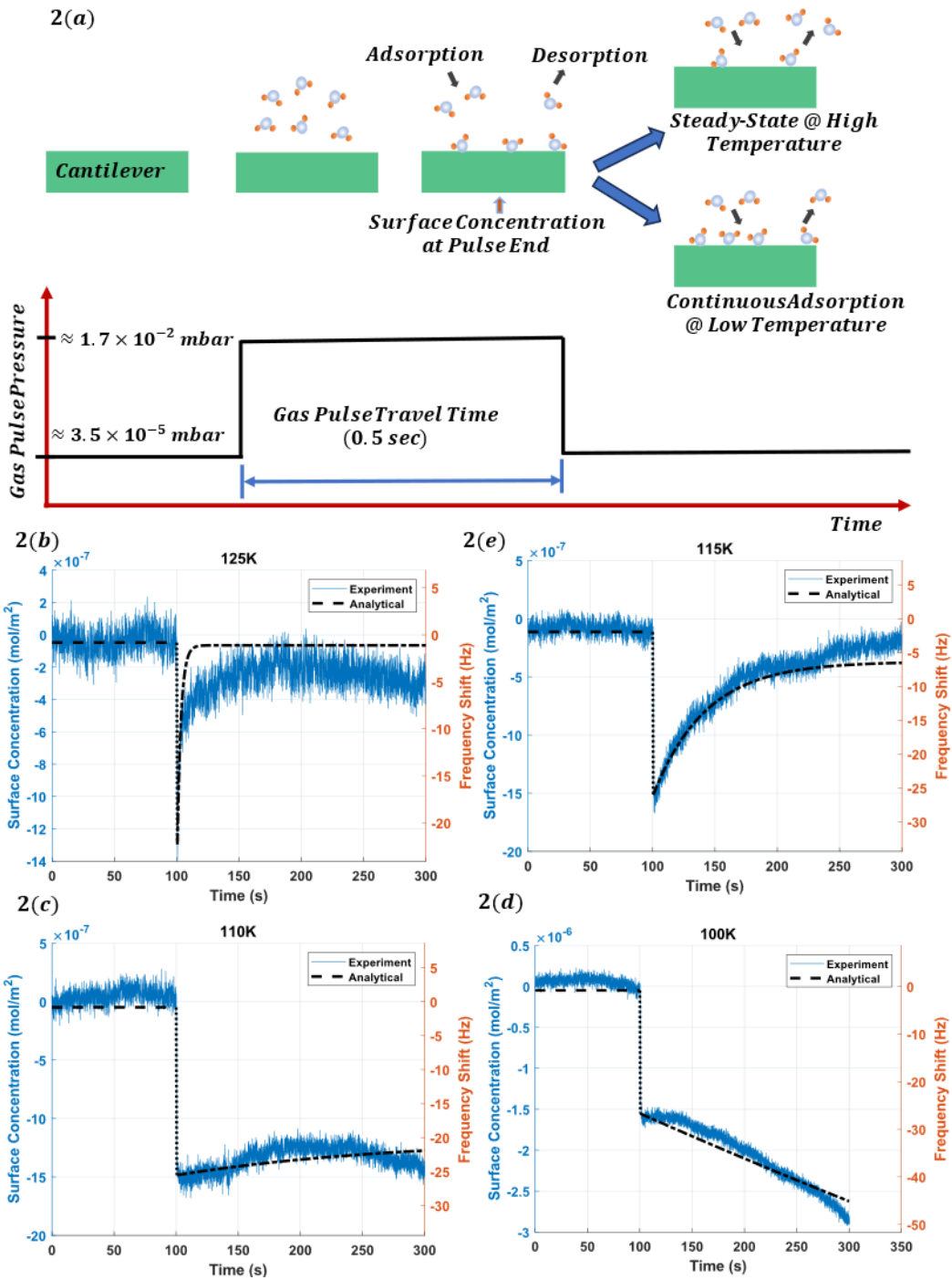

*Figure 2 : (a) Illustration of the gas pulse sequence and the adsorption-desorption process on the device surface. 2(b)-(e) Adsorption-desorption responses with time at different temperatures, when UHP helium gas is pulsed into the chamber. The black dashed lines are obtained using the analytical model, assuming UHP helium with 18 ppm moisture content.*

frequency shift only below ~140 K in this helium ambient. The desorption rate increases with temperatures [19]. Hence, above the threshold temperature of 140 K, the residence time of adsorbed species is much smaller than the device's response time, so the observed frequency shift is negligible. We observe a direct correlation between the threshold temperature and the amount of trace moisture in the UHP gas. Argon with a lower moisture concentration of 9 ppm shows a frequency shift only below 125 K.

Apart from the magnitude of the frequency shift, which correlates with the moisture concentration in these gases, the device's response to both gas pulses is similar. Above 109 K, the device shows a momentary dip in resonance frequency when gas is pulsed into the chamber. Thereafter the resonance frequency gradually recovers, settling at a slightly lower value than the original frequency. We assert that the evolution of frequency shift mirrors the kinetics of water physisorption on the microcantilever. Below 109 K, there is a continuous decrease in the resonance frequency, beyond the initial dip, probably due to continuous physisorption of moisture onto the cantilever. This transition temperature will also depend on the chamber pressure.

We use analytical and finite element method (FEM) simulation (COMSOL) to understand these observations. The adsorption and desorption of gas and moisture molecules on the cantilever can be represented by the reaction:

$$Molecules_{vapour} + Vacant\ sites_{surface} \leftrightarrow Molecules_{surface} \qquad 1$$

The term $Molecules_{vapour}$ and $Molecules_{surface}$ refer to the concentration of vapor and adsorbed molecules, respectively. The molecules can be target gas molecules or trace moisture. Before the gas pulse, both vapor and adsorbed concentration are zero, i.e. $c(0) = 0$. The concentration increases rapidly when the gas pulse is applied. The gas concentration "χ" ppm, is related to the partial pressures by equation 2, where P$_{Chamber}$ is the measured chamber pressure.

$$p_{H_2O} = \frac{\chi}{10^6} P_{Chamber}; \quad p_{Gas} = 1 - p_{H_2O} \qquad 2$$

The adsorption rate of molecules on the cantilever's surface can then be described by:

$$R_{ads} = \frac{p}{\sqrt{2\pi M \, k_B T_{Chamber}}} * (1-\phi) * e^{-\frac{E_{ads}}{k_B T_{Cantilever}}} = k_{ads} * (1-\phi) \qquad 3$$

Here, $\frac{p}{\sqrt{2\pi M \, k_B T_{Chamber}}}$ (Hertz-Knudsen equation [20]) represents the flux of molecules at the surface for a partial pressure of $p$ of either the target gas or moisture. $E_{ads}$ and $M$ are the adsorption energy and molecular weight, respectively, $k_B$ is the Boltzmann constant and $T_{Chamber} \sim 300K$ and $T_{Cantilever}$ refers to the temperature of the chamber and the cantilever, respectively. $k_{ads}$ represents the adsorption rate constant. If $c_{sites} \sim 10^{19} \frac{sites}{m^2}$ is the total number of available sites and $c(t)$ is the number of molecules adsorbed on the surface, then $\phi(t) = \frac{c(t)}{c_{sites}}$ represents the fraction of "adsorbed" sites and (1-$\phi$) is the "open-site" fraction. The exponential term in Equation 3 represents the probability of a successful attachment of molecules on the surface. For physisorption, $E_{ads} \approx 0$ for both target gas and water molecules [19].

The desorption rate of molecules ($R_{des}$) from the cantilever's surface to the vapor phase can be described by:

$$R_{des} = \nu * e^{-\frac{E_{des}}{k_B T_{Cantilever}}} * c(t) = k_{des} * c(t) \qquad 4$$

$$k_{des} = \nu * e^{-\frac{E_{des}}{k_B T_{Cantilever}}} \qquad 5$$

Here, $\nu$ is the attempt frequency of the molecules on the surface, which is given by $k_B T_{Cantilever}/h$, where $h$ is the Planck's constant. $E_{des}$ is the energy barrier for the desorption of the molecules from the surface. The exponent in the equation 4 corresponds to the probability of a successful desorption event. Thus, the term $\nu * e^{-\frac{E_{des}}{k_B T_{Cantilever}}}$ signifies the frequency of successful detachment of molecules from the surface. The mean residence time ($\tau$) of the molecules is inversely proportional to desorption rate constant, $k_{des}$.

Therefore, the total rate of adsorption-desorption can be calculated as:

$$\frac{dc(t)}{dt} = R_{ads} - R_{des} = k_{ads} * (1-\phi) - k_{des} * c(t) \qquad 6$$

Initially, the desorption rate, $R_{des}$ will be zero because the surface concentration of gas/water molecules (c(t)) is zero. On the other hand, the adsorption rate, $R_{ads}$ will be at its maximum because the surface concentration of molecules, i.e. $\phi(t) = c(t)/c_{sites}$ is zero. Over time, molecules adsorb onto the cantilever's surface, leading to an increase in $c(t)$, and subsequently raising $R_{des}$ and reducing $R_{ads}$. Eventually, the system reaches a steady-state where adsorption and desorption rates are equal and surface concentration is constant. As expected, the above equations show that higher vapour pressure ($p_{Gas}/p_{H_2O}$) and lower temperatures contribute to a higher steady-state concentration on the surface.

These equations can be solved for target gases and water molecules. The gases stay in the chamber for approximately 0.5 seconds. For the simulations, we assume a highly simplified gas pulse which increases the chamber pressure to a constant value for 0.5 sec and then transition sharply back to zero due to pumping.

Furthermore, the analytical solution can be used to fit the experimental results to obtain $E_{des,H_2O}$ and the concentration of gas (dashed lines in Figs. 2(b)-2(e)). The experimental fit indicates that $E_{des,H_2O}$ is approximately 0.31eV to 0.32 eV and the concentration is found to be around 18 ppm. The obtained desorption energy is well within the range of previously reported values for water molecules [21]. And the moisture concentration matches well with the CDRS measurements on the UHP helium gas (Figure. S4 (SI)).

We performed a similar analysis for helium gas itself. However, for desorption energies ranging from 0.05 eV to 0.20 eV [21-25], which are typical for inert gases, the analytical solution does not fit the experimental results and indicates permanent adsorption at all desorption energies and temperatures (Figure. S6 SI). Additionally, the surface concentration of water molecules is approximately 6 to 8 orders of magnitude higher than that of helium or argon molecules (SI Figure. S7) considering $E_{des,He/Ar} = 0.05\ eV$. This suggests that the moisture content in these UHP gases predominantly influences the microcantilever's frequency response and frequency shift. These findings further support the model of moisture adsorption as the primary driver of frequency shifts, with minimal contribution from the intended UHP gas.

If the observed frequency shift is indeed the result of moisture adsorption, we expect the following two effects: a) The frequency shift will vary with the moisture concentration in the target gases, not with the gases themselves and b) The transition from temporary to permanent adsorption would happen at the same temperature irrespective of the target gas. We performed measurements at different temperatures to investigate these effects. Figure 3 shows the experimentally observed frequency response of the microcantilever at different temperatures using gas pulses of helium and argon. We found that the frequency shifts correspond to the moisture content in helium and argon, and the magnitude of the frequency shift depends on the moisture concentration. For the argon gas pulse, the simulated frequency response slightly deviated from the experimental one at temperature above ≈ 110 $K$ due to a slight difference in the chamber pressure at the end of the pulse (cyan-colored line). Figure 2(b-e) and SI Figure S5 show that the transition from temporary to permanent adsorption occurs at approximately 109 K for both He and argon. These results further prove that the adsorbed species is moisture, and adsorption is independent of the target gas.

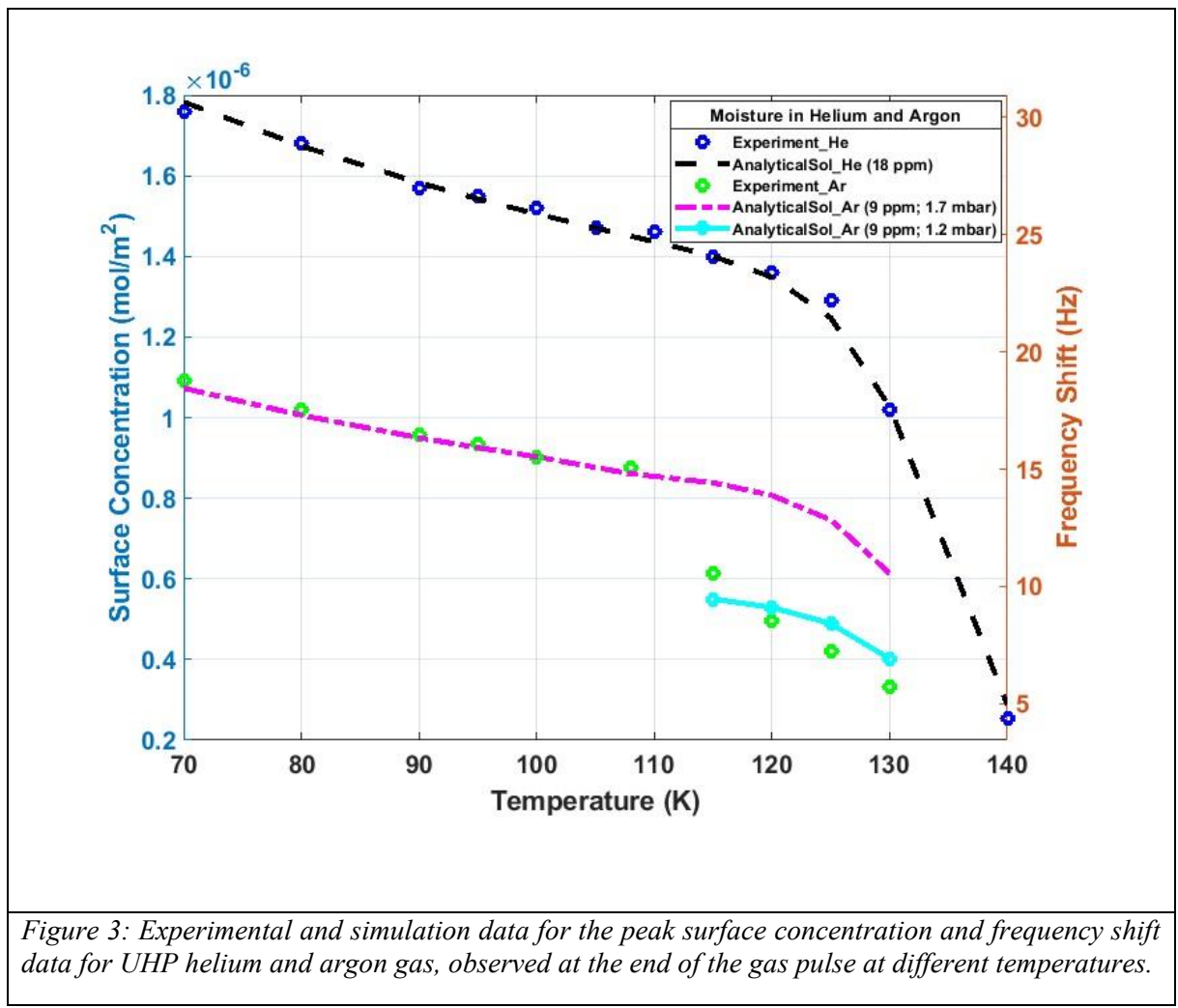

*Figure 3: Experimental and simulation data for the peak surface concentration and frequency shift data for UHP helium and argon gas, observed at the end of the gas pulse at different temperatures.*

The frequency response of the microcantilever across various temperature ranges can be inferred from the data illustrated in Figure 4. Figure 4(a) presents simulation data of the surface concentration of water molecules at the end of the pulse (black line). Also, shown are the steady-state surface concentrations ($[H_2O]_{Steady-state,}$) at pressures of $3.4 \times 10^{-5}$ mbar (blue line) and $1.7 \times 10^{-2}$ mbar (red line). These lines merge at the lower temperatures as the total number of available sites will be filled under those conditions. The temporal frequency response depends on the interplay between these values. The time to achieve steady-state surface concentration is calculated using equations 6. It is important to note that, at around 109 K, a convergence of the blue and black lines is observed. This indicates that above 109 K, the steady-state concentrations at pressure $3.5 \times 10^{-5}$ mBar are lower than the surface concentration at the pulse end. As a result, temporary adsorption of molecules occurs, which eventually desorbs over time, reaching a steady-state surface concentration. As we lower the temperature, this recovery takes longer (see Figures 2(b) and 2(c)). Below 109 K, the surface concentration at the pulse end is lower than the steady-state concentration, resulting in continuous adsorption.

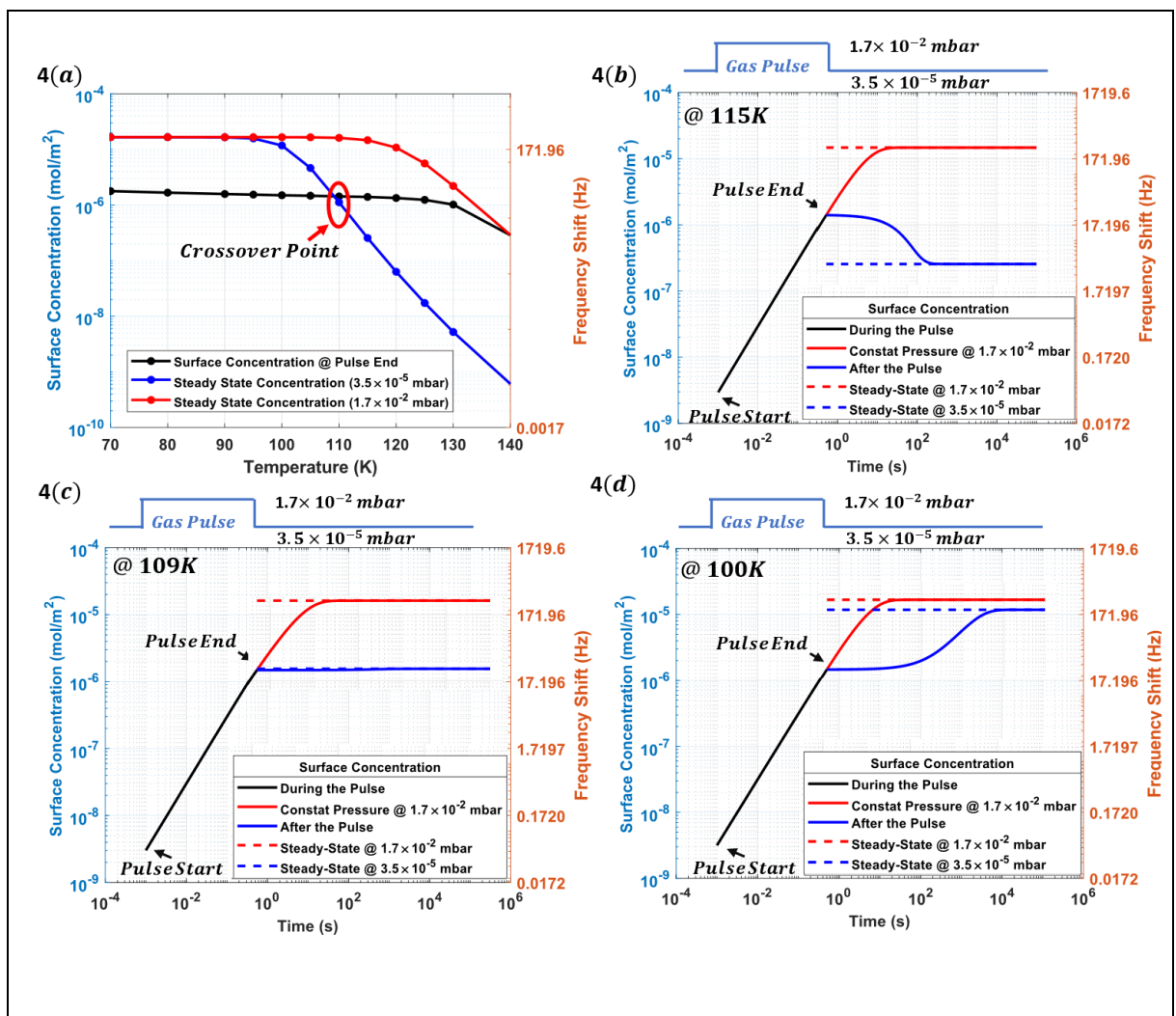

Figure 4: (a) Simulated surface concentration at pulse end and steady-state surface concentrations of water molecules on the microcantilever. The convergence of the surface concentration at pulse end and the steady state surface concentration (@$3.5 \times 10^{-5}$ mbar) at 109 K suggests a shift from transient to permanent water molecule adsorption below this temperature. Figures. 4(b)-(d) show the temporal variation of surface concentration on the device surface at different temperatures, as chamber pressure cycles from $3.5 \times 10^{-5}$ mbar to $1.7 \times 10^{-2}$ mbar and back, upon pulse completion.

Figures 4(b-d) illustrate these phenomena differently across various temperatures. The black line represents the adsorption-desorption dynamics during the gas pulse's entry into the system, assuming a constant chamber pressure of about $1.7 \times 10^{-2}$ mBar. After 0.5 seconds, the chamber pressure decreases to roughly $3.5 \times 10^{-5}$ mBar, with the blue curve representing the actual device response under these conditions. The solid red curve depicts the hypothetical device response if the chamber pressure remained at $1.7 \times 10^{-2}$ mBar. Dashed red and blue lines indicate the steady-state concentrations of water molecules at pressures of approximately $1.7 \times 10^{-2}$ mBar and $3.5 \times 10^{-5}$ mBar, respectively. At each temperature, once the gas pulse is introduced, the surface concentration increases. If the chamber pressure is maintained at $1.7 \times 10^{-2}$ mBar, the concentration would track along the solid red curve and eventually reach the dashed red line. However, once the pulse ends, the chamber pressure decreases to $3.5 \times 10^{-5}$ mBar causes the concentration to follow the solid blue line and ultimately reach the dashed blue line after reaching a steady state. Importantly, these steady-state surface concentrations depend on temperature. For temperatures above 109K, as gas pulses enter the chamber, surface concentration increases, initially corresponding to a decrease in resonance frequency or an increase in frequency shift (See Figure 2(b-d)). Since the chamber is being continuously pumped, the gas is quickly pumped out, and the surface concentration doesn't reach the steady state concentration corresponding to $1.7 \times 10^{-2}$ mBar pressure (dashed red line) at the end of the pulse. If the gas pulse duration were longer, the surface concentration would have reached the dashed red line. Instead, at the end of the pulse, it starts approaching the lower surface concentration that is expected at $3.5 \times 10^{-5}$ mBar pressure. This corresponds to the desorption process and leads to recovery in the resonance frequency. As we lower the temperature, this recovery takes longer (see Figure 4(b) and (c)).

At around 109 K, the surface concentration at the end of the pulse is very close to the surface concentration in the steady state (Figure 4(c)). Therefore, the frequency shift becomes permanent. This corresponds to a step-like frequency shift, as observed in Figure 2(d) (and Figure S5(c) SI). Below 109 K, the steady-state concentration is higher than the surface concentration at the end of the pulse. This results in adsorption even after the pulse has ended, and we observe a frequency shift that is continuous with time.

Figure 5 (a)-(c) shows simulation results of the device's behaviour at 115 K, with varying moisture concentrations in the gas pulse. The simulations reveal that altering the moisture concentration affects both the initial surface concentration at the end of the pulse and the steady-state concentrations. Despite these changes, the time taken to reach the steady-state concentration remains constant due to an increase in the rates proportional to the concentration. Figure 5(d) illustrates the initial and steady-state surface concentrations of the devices under different moisture concentrations in the pulse. Regardless of the moisture concentration in the gas pulse, the initial surface concentration and the steady-state concentration at $3.4 \times 10^{-5}$ mbar converges at the same temperature. In terms of frequency shift, this implies that the crossover temperature from temporary to permanent adsorption will be the same temperature if the gas adsorbing onto the cantilever is the same. This is consistent with experimental results obtained using UHP helium and argon pulses, as detailed in Figure 2(b) to (e) and Figure S5 (SI). Even though two different gases are introduced, the frequency shift observed matches well with the trace concentration of moisture in these UHP gases and the transition to permanent adsorption occurs at the same temperature.

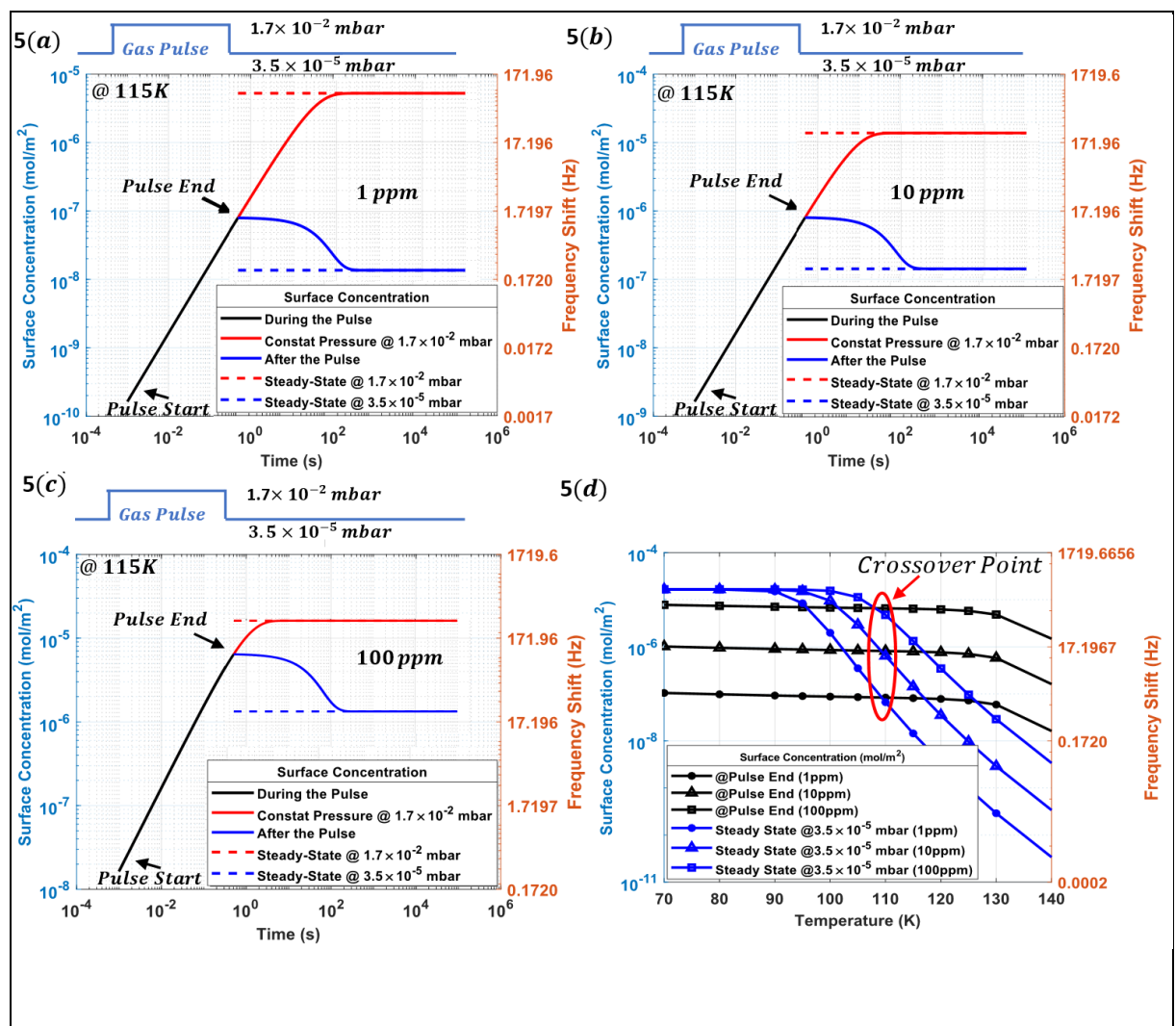

*Figure 5: (a)-(c) Simulation data for the changes in surface concentration on the device at 115 K with varying moisture levels in the gas pulse, as chamber pressure cycles from $3.5 \times 10^{-5}$ mbar to $1.7 \times 10^{-2}$ mbar and back, upon pulse completion. (d) Shows both initial (at pulse end) and steady-state surface concentrations across various temperatures. The intersection of these concentration curves at 109 K suggests a shift from transient to permanent water molecule adsorption happens below this temperature, regardless of the moisture level in the gas pulse.*

Figure S8 in the SI presents a similar analysis of the impact of varying chamber pressure at the pulse end. Altering the chamber pressure from 0.1 Pa to 10 Pa modifies the initial surface concentration at the end of the pulse. However, the chamber pressure influences the transition temperature at which the device response shifts from a momentary to a permanent change (SI Figure S8(d)). The simulation in SI Figure S8(a-c) illustrates these results for the adsorption-desorption response at 115 K. At low chamber pressure (0.1 Pa), the transition temperature is approximately 120 K. In this case, the initial surface concentration at the pulse end is lower than the steady-state surface concentration, resulting in a permanent shift due to continuous adsorption of molecules. Conversely, at higher chamber pressures (1 Pa and 10 Pa), the transition temperatures occur at lower values, and the resonance frequency recovers after adsorption. We have also modelled the system using finite element method (FEM) in COMSOL Multiphysics to incorporate the effect of diffusion of the gases. Figure. S9 in the SI illustrates the surface concentration of water molecules at the end of the pulse across various temperatures, as determined by both FEM and analytical simulation. The data from the FEM simulation aligns precisely with the analytical results, suggesting that incorporating diffusion kinetics does not impact the surface concentration of molecules adsorbing onto the device's surface. These noteworthy results, obtained from both analytical and FEM simulations along with experimental data, provide a comprehensive insight into the frequency response of the microcantilever during the physisorption of the gases. They highlight a consistent response in frequency shifts across various gases and elucidate how factors such as moisture content, gas pulse width, and temperature variations influence the magnitude of these frequency shifts.

**CONCLUSION**

In summary, we confirm that the frequency shifts in uncoated silicon nitride microcantilevers mainly come from the physisorption of trace moisture in UHP gases. We proposed a model that consistently explains the data for different gases, like helium and argon, and under various conditions, including different moisture levels and temperatures. We have observed that beyond a specific temperature, which depends on the chamber pressure, the frequency shift becomes permanent. These findings are further validated through simulations, which show good agreement with experimental data and analytical solutions, thereby supporting our conclusions.

This study improves our understanding of how microcantilevers interact with gases at the nano-level. It is important for measuring physio adsorption using micro and nanomechanical resonators. Any such measurements need to account for the desorption energy levels and gas concentrations present. This has implications for various fields, including environmental monitoring, industrial safety, and healthcare diagnostics, where detecting minute changes in gas composition is crucial.

**ACKNOWLEDGEMENT**

We gratefully acknowledge Varun Adiga, Yash Bajpai, and Prof. Praveen Ramamurthy for their assistance with the CRDS measurements conducted in their laboratory.

# Physisorption on Nanomechanical Resonators: The Overlooked Influence of Trace Moisture

Hemant Kumar Verma[1], Suman Kumar Mandal[1], Darkasha Khan[1], Faizan Tariq Beigh[1], Manoj Kandpal[2], Jaspreet Singh[2], Sushobhan Avasthi[1], Srinivasan Raghavan[1] and Akshay Naik[1*]

1. **Device Fabrication**

The fabrication process involves the utilization of a highly doped silicon substrate, which also acts as a back gate for the electrostatic actuation of the device. Subsequently, an approximately 1um thick layer of undoped silicate glass (USG) is deposited through Low-Pressure Chemical Vapor Deposition (LPCVD). This particular layer serves as a sacrificial layer in the fabrication process. In the subsequent stage, a structural layer of silicon nitride, measuring approximately 300nm in thickness, is deposited onto the stack of silicon and USG layers using LPCVD. The microcantilever itself, with dimensions of roughly 30um in length, 5um in width, and 300nm in thickness, is then defined through a process of reactive ion etching (RIE). Following this step, the piezo resistors made of metal (Cr/Au with thicknesses of approximately 5nm/25nm) and metal pads (Cr/Au with thicknesses of approximately 10nm/60nm) are patterned onto the structure through optical lithography and deposited using an e-beam evaporator. Subsequently, a metal lift-off process is employed to define the desired features accurately. Finally, to release the microcantilever from the substrate, Buffered Hydrofluoric Acid (BHF) is used in a wet etching process. To prevent the collapse of the cantilever due to surface tension, a Critical Point Dryer (CPD) is utilized. Figure 1 in the paper displays images of the fabricated microcantilevers [1][2].

2. **Measurement Circuit**

The experiments were conducted utilizing a silicon nitride microcantilever in a vacuum environment with a pressure level of $10^{-5}$ millibars. Supplementary Figure 1 illustrates a simplified schematic of the circuit employed for these measurements. The methodology employed for measurement involved the use of the electrical homodyne technique, along with electrostatic actuation and metal piezo resistors for transducing the microcantilever's motion [1][2][3]. This setup featured a cantilever equipped with a metal piezoresistor on its upper surface and the highly doped silicon wafer serving as the back-gate electrode for electrostatic actuation. Additionally, three resistors of the same material and dimensions were positioned on the fixed substrate to form a Wheatstone bridge circuit.

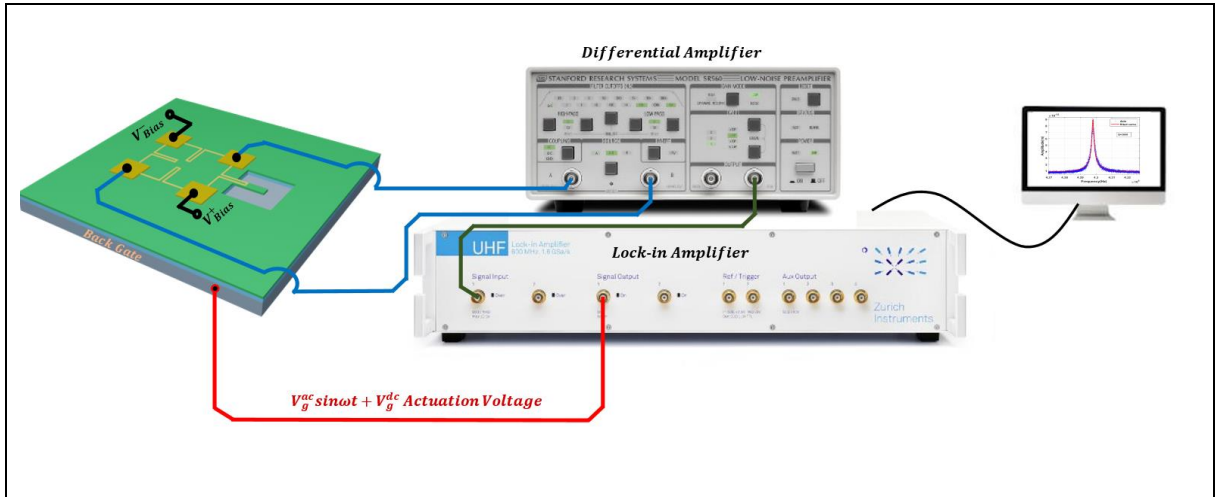

*Figure S1: Schematic showing the homodyne electrostatic actuation and piezoresistive detection scheme for the microcantilever. We applied a $V_{Bias}=\pm 15\ mV$ at two opposite nodes of the Wheatstone bridge circuit and the output taken from the other two opposite node with SRS560 preamplifier and Zurich Instrument UHF lock-in amplifier. The microcantilever is electrostatically actuated by applying AC+DC voltages between the top metal and the backside silicon wafer.*

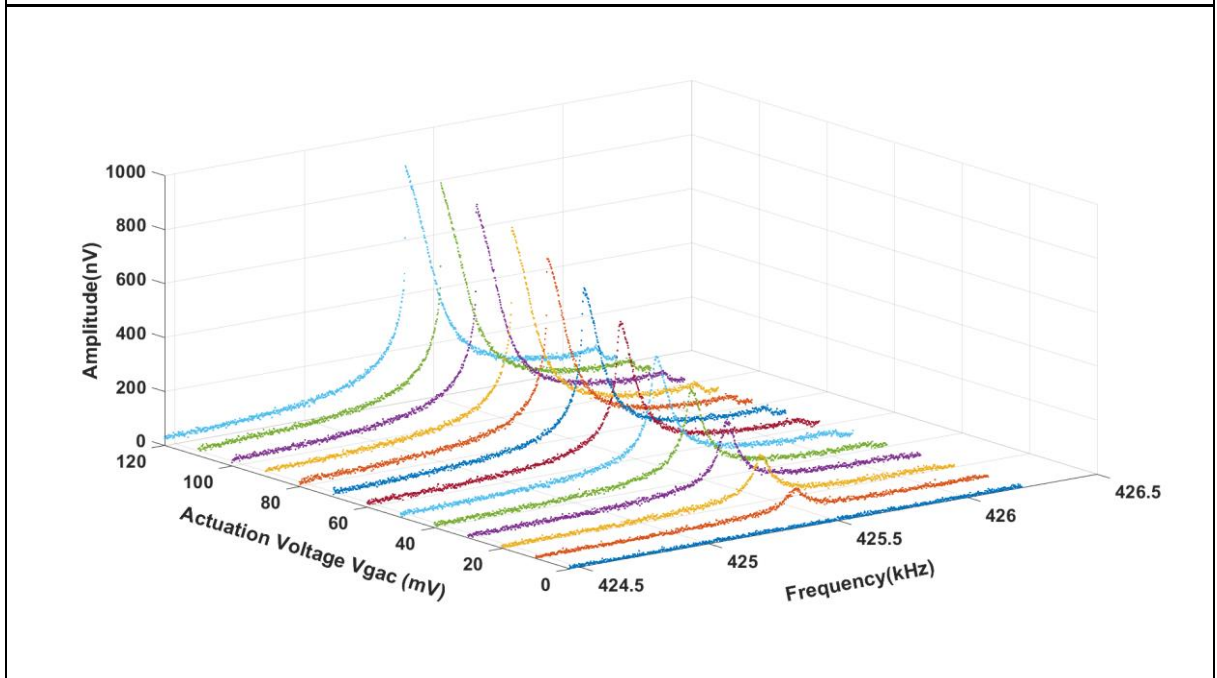

*Figure S2: Amplitude versus frequency response of the $Si_3N_4$ microcantilever at different actuation forces. In this context, $V_g^{dc}$ is held constant at 3V while $V_g^{ac}$ is varies. We observed the non-linearity in the device response after $V_g^{ac} = 70\ mV$ at room temperature.*

The implementation of this Wheatstone bridge circuit played a crucial role in enhancing the signal-to-noise ratio (SNR) and compensating for temperature fluctuations. To electrostatically actuate the devices, a small AC signal ($V_{gac}sin\omega t$) and a substantial DC voltage ($V_{gdc}$) were supplied from a lock-in amplifier (Zurich Instruments UHFLI) between the back gate and the metal piezoresistor on top of the cantilever. Furthermore, a DC bias of approximately 30 millivolts ($V_{Bias} \approx 30 mV$) was applied to the bridge circuit. As the cantilever underwent motion, the resistance of the piezoresistor underwent changes, and these variations in voltage were detectable via the lock-in amplifier.

### 3. Moisture content in the UHP gases

The moisture presents in the UHP gases is measured with cavity ring-down spectroscopy (CRDS) instrument. CRDS is a highly sensitive analytical technique employed to measure trace gas concentrations, making it well-suited for determining moisture content in ultra-high purity gas cylinders. In CRDS, a tuneable laser emits light into an optical cavity filled with the sample gas. The gas molecules absorb specific wavelengths of the laser light, reducing the light intensity inside the cavity. CRDS calculates moisture concentration by measuring the time it takes for the laser light to decay (ring down) inside the cavity, with shorter ring-down times corresponding to higher moisture levels. By calibrating the system with known standards and analyzing the ring-down time, CRDS provides precise measurements of moisture content in ultra-high purity gases, offering critical insights into gas quality for various applications. In Figure 4, the CRDS data illustrates the moisture levels in ultra-high purity (UHP) helium and argon gases. Initially, the moisture content was notably elevated due to insufficient purging when connecting the gases to the CRDS system. Consequently, as time progressed, the moisture concentration reached saturation, providing us with the ultimate measurement of moisture content present within these gas cylinders.

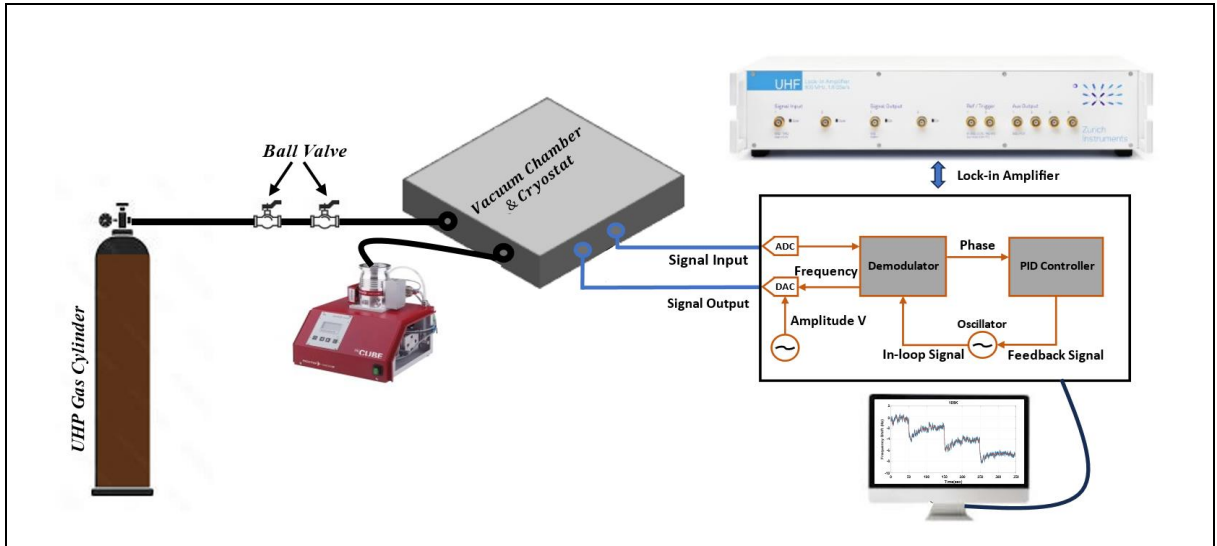

*Figure S3: Schematic of the measurement setup to study the temperature-dependent frequency response of the microcantilever and quantify the moisture present in the UHP gases in real-time with PLL circuits.*

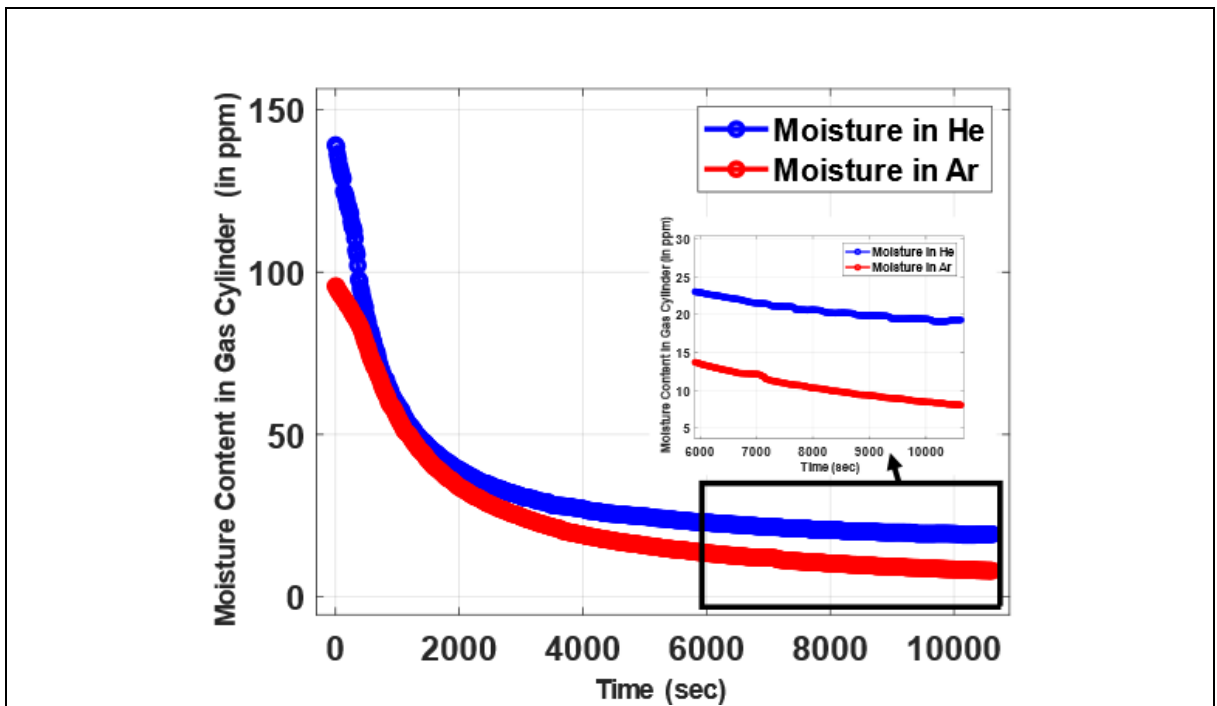

*Figure S4: CRDS measurements data for the moisture levels in the UHP He and Ar gases. Initially, higher moisture content was recorded, attributed to inadequate purging during the connection of the gas cylinders to the CRDS systems. The inset displays approximately 18 ppm and 9 ppm of moisture present in the UHP helium and argon cylinders, respectively, following saturation.*

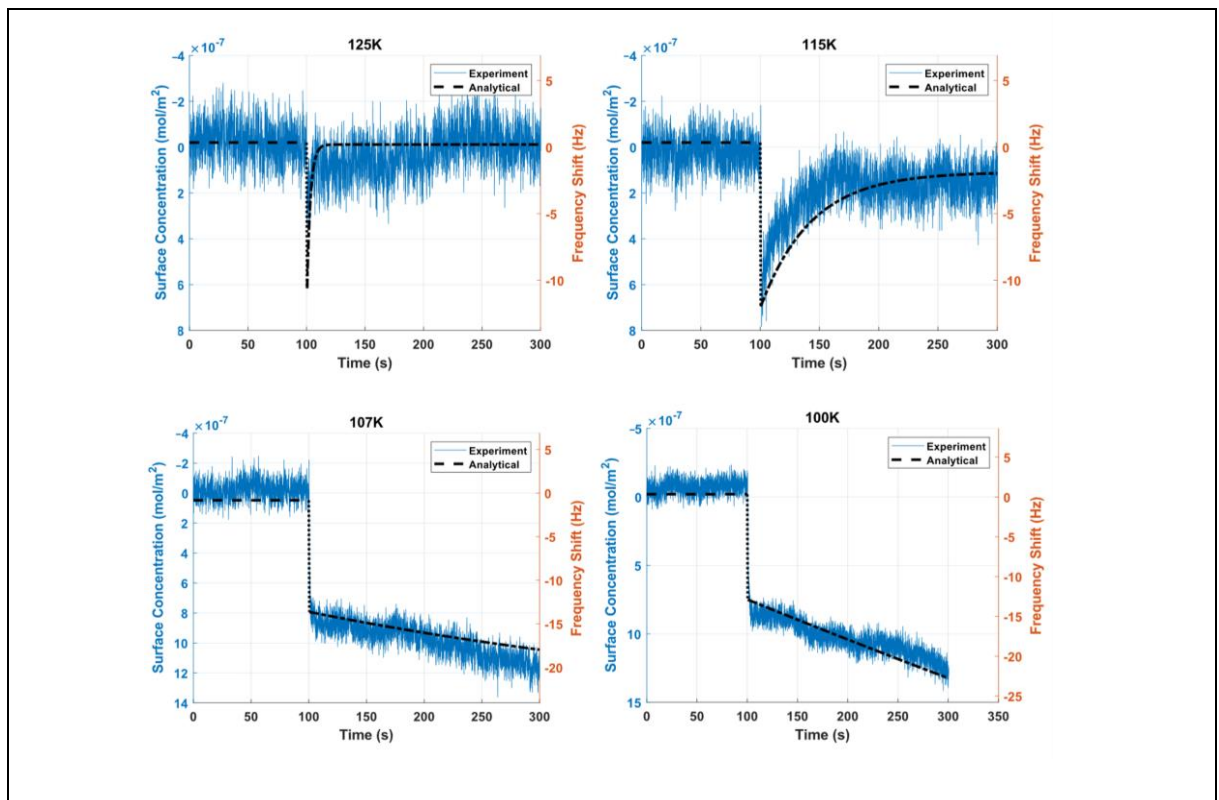

*Figure S5: The experimental results for UHP argon gas, displaying the adsorption-desorption behavior on the device surface at various temperatures. The analytical model, overlapped on the experimental data, indicates a moisture level of around 9 ppm. Remarkably, the device's response transitions from a temporary to a permanent shift at 109 K, identical to that observed with helium. gas, and the energy barriers are found to be exactly the same.*

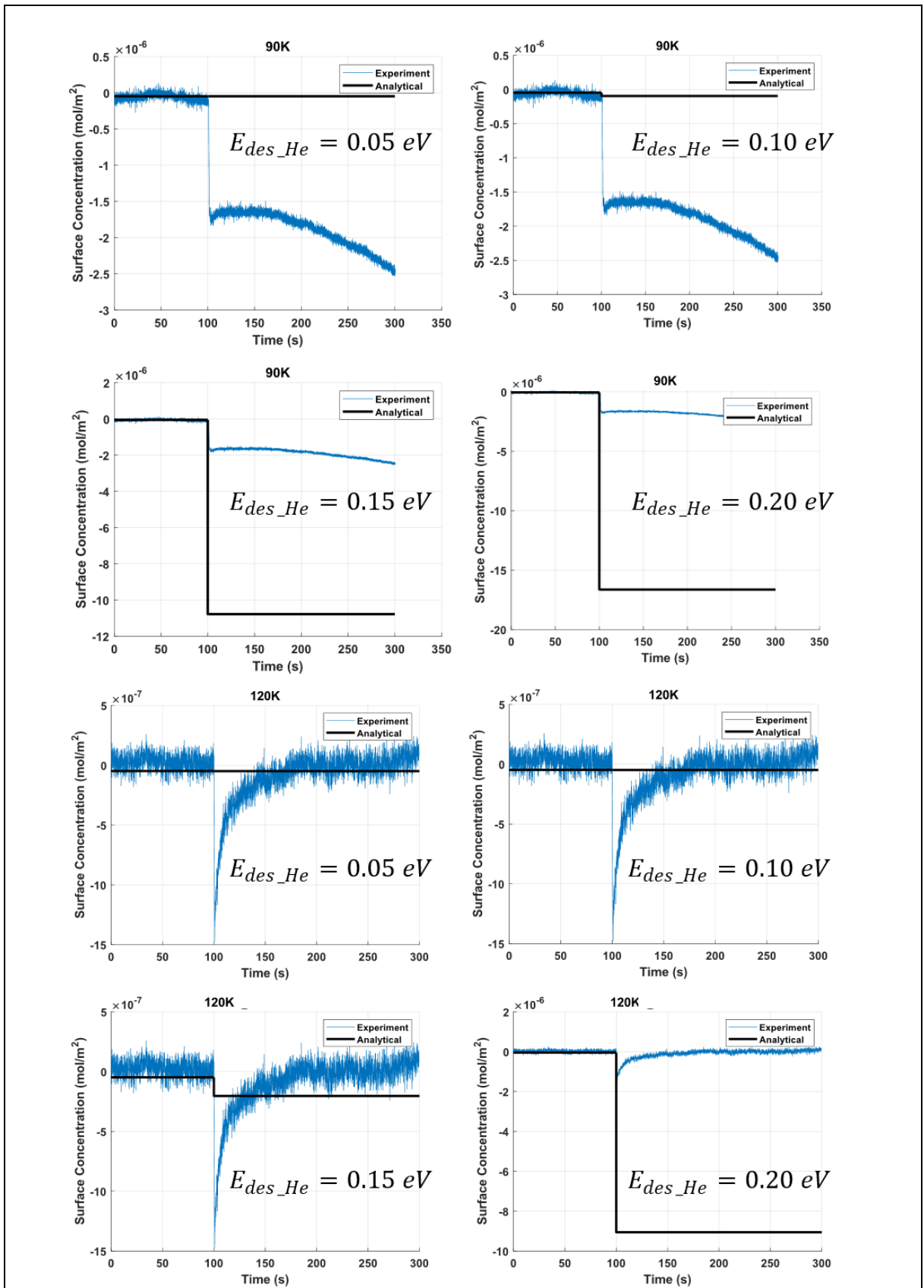

*Figure S6: Fitting analytical frequency response considering Helium molecules as the adsorbent on the experimental data with $E_{des,He}$ ranging from 0.05 eV to 0.20 eV at 90K and 120K.*

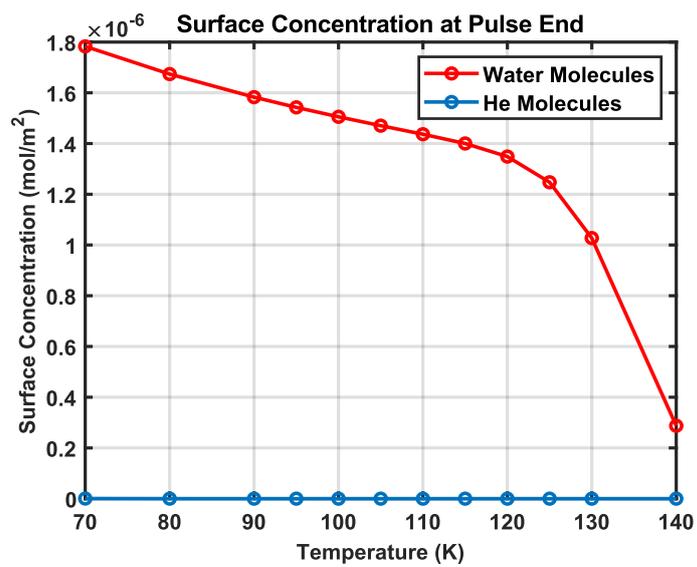

*Figure S7: Simulation data depicting the surface concentrations of water and helium molecules at the pulse end across various temperatures considering $E_{des}$ of 0.32eV and 0.05 eV for water and Helium molecules, respectively.*

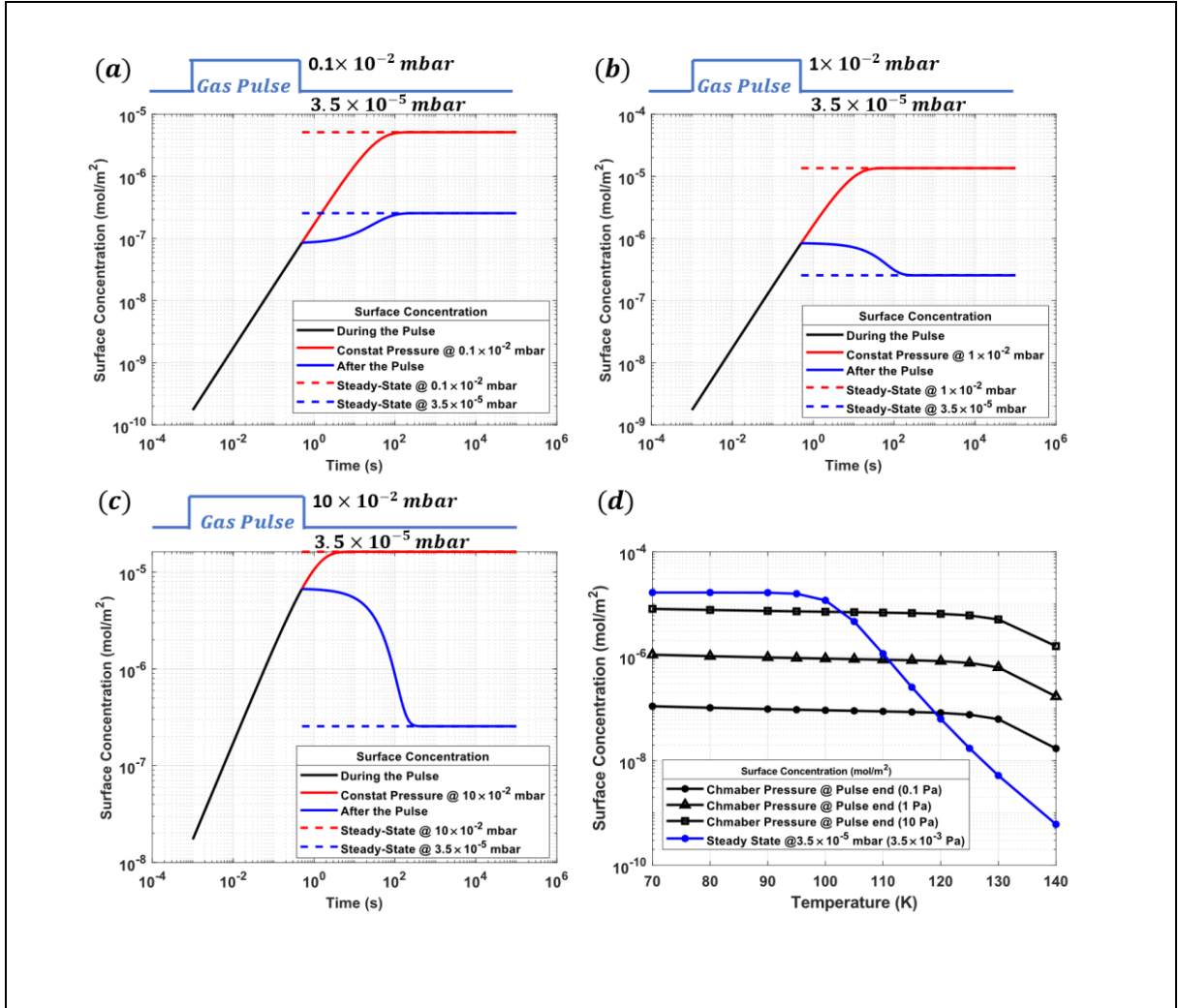

*Figure S8:(a)-(c) Simulation data for the surface concentration variations on the device at 115 K with different chamber pressures at pulse end. (1 Pa=0.01 mbar) (d) shows both initial (at pulse end) and steady-state surface concentrations at varying temperatures, with the intersection points at different temperatures indicating the transition temperatures for the device's shift from temporary to permanent response*

### 4. Analytical versus FEA Simulation

We have also modelled the system using finite element method (FEM) in COMSOL Multiphysics to incorporate the effect of diffusion of the gases. The $Gas|H_2O$ molecules move from the vapour to the cantilever's surface by diffusion through a boundary layer formed at the vapour-surface interface. The diffusion kinetics through the boundary layer follows Fick's second law of diffusion. The diffusion coefficient of gas molecules is usually in the $10^{-5}$ to $10^{-6}\ m^2/sec$ range [4]. We assume a diffusion coefficient of $10^{-5}\ m^2/sec$ for all the gases in this model. The transport of dilute species interface under the chemical transport branch in the COMSOL Multiphysics® software is considered to calculate the local $p_{Gas}$ and $p_{H_2O}$ near the cantilever's surface. This local $p_{Gas}|p_{H_2O}$ is then considered in Equation 5.3 to calculate the rate of adsorption of $Gas|H_2O$ molecules. To solve the kinetics of the adsorption and desorption processes, the general form boundary PDE interface is used in the COMSOL Multiphysics® software under the mathematics branch. The heat transfer through fluids module is utilized to specify the temperatures of both the chamber and the microcantilever. The benefit of using the FEM simulation is to utilize the diffusion kinetics through the boundary layer. Figure. 5.14 in illustrates the surface concentration of water molecules at the end of the pulse across various temperatures, as determined by both FEM and analytical simulation. The data from the FEM simulation aligns precisely with the analytical results, suggesting that incorporating diffusion kinetics does not impact the surface concentration of molecules adsorbing onto the device's surface.

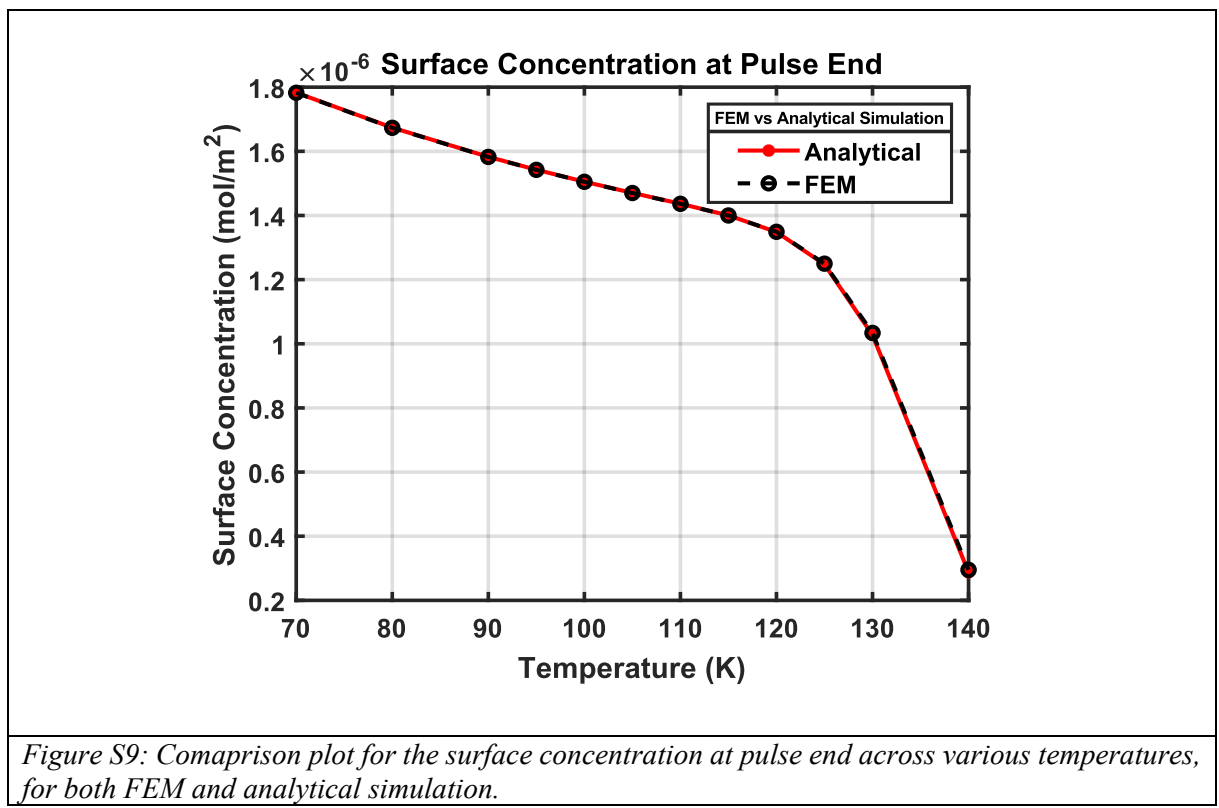

*Figure S9: Comaprison plot for the surface concentration at pulse end across various temperatures, for both FEM and analytical simulation.*

## References

[1] H. K. Verma, D. Khan, M. Kandpal, S. N. Behra, J. Singh, and A. Naik, "Silicon Nitride Microcantilever-Based Temperature Sensors," APSCON 2023 - IEEE Appl. Sens. Conf. Symp. Proc., pp. 1–3, 2023.

[2] H. K. Verma, F. T. Beigh, D. Khan, M. Kandpal, S. N. Behra, J. Singh, and A. Naik, "Frequency response of uncoated-microcantilevers to gas flow at different